\documentclass[]{JFM-FLM_Au}

\usepackage{graphicx}
\usepackage{amsmath}
\usepackage{bm}
\usepackage{siunitx}
\usepackage{subcaption}
\usepackage{hyperref}
\usepackage{booktabs}
\usepackage{amssymb}
\usepackage{cleveref}

\lefttitle{S. P. Kalathoor and J. C. Oefelein}
\righttitle{Filtering and entropy transport in a reacting shear layer}

\title{Filtering effects on entropy transport and entropy-production structure in a supersonic reacting shear layer}

\author{Sriram P. Kalathoor\aff{1}\,
  \and Joseph C. Oefelein\aff{1}}

\affiliation{\aff{1}Daniel Guggenheim School of Aerospace Engineering, Georgia Institute of Technology, Atlanta, GA 30332, USA}

\corresau{Sriram P. Kalathoor, \email{sriram@gatech.edu}}

\newcommand{\figpath}{figures/clean}

\begin{document}
\maketitle

\begin{abstract}
Spatial filtering is examined in time-resolved mid-plane DNS fields of a supersonic reacting shear layer using a sequence of box filters. The analysis tracks a nondimensional entropy-like scalar $s^\ast$, its in-plane material derivative $D s^\ast / D t^\ast$, a residual $\Pi_s^\ast$, and viscous and conductive entropy-production diagnostics, $\sigma_\mu^\ast$ and $\sigma_k^\ast$. Filtering changes $s^\ast$ only weakly, attenuates $D s^\ast / D t^\ast$ and the strongest tails of $\sigma_\mu^\ast$ and $\sigma_k^\ast$, but broadens the residual distribution and increases the residual RMS with filter width. The residual remains concentrated in the layer core that carries the largest mechanical and thermal activity. Conditional statistics show that $|\Pi_s^\ast|$ rises with both entropy-production intensity and entropy-gradient strength. Spectral and structural diagnostics show that increasing filter width removes high-wavenumber content and simplifies the geometry of the high-$|\Pi_s^\ast|$ sets. Coarser filtering therefore increasingly distorts entropy transport preferentially through the most dynamically and thermally active structures, rather than uniformly across the plane.
\end{abstract}

\section{Introduction}
Entropy-production diagnostics provide a compact view of irreversibility in compressible reacting flows, where viscous deformation, heat conduction, species or scalar transport, and chemical conversion enter the local entropy balance \citep{DeGrootMazur1962}. In turbulent shear layers these processes are strongly localized because entrainment, interfacial sharpening, and reaction occupy only part of the mixed region \citep{BrownRoshko1974,Breidenthal1981,BroadwellBreidenthal1982,HermansonDimotakis1989,PantanoSarkarWilliams2002,PantanoSarkarWilliams2003}. Compressibility further modifies the large structures, gradient field, and thermochemical response \citep{PapamoschouRoshko1988,SandhamReynolds1991,PantanoSarkar2002,LivescuJaberiMadnia2002,Lele1994,MahleEtAl2007,JahanbakhshiMadnia2016,JahanbakhshiMadnia2018}.

The same localization makes filtering especially consequential. In large-eddy descriptions, filtering suppresses the small-scale content that carries much of the local gradient intensity and dissipation, so the strongest errors should be expected where the sharpest structures are concentrated \citep{Pope2000,Sagaut2006,MeneveauKatz2000,FoysiSarkar2010}. For reacting shear layers, the relevant questions therefore have spatial as well as statistical implications. Where does the entropy-transport residual localize, how does its amplitude change with filter width, and does it remain tied to the mechanically and thermally active structures of the unfiltered layer?

Those questions are physically meaningful because prior studies of compressible and reacting mixing layers show that the layer core carries a disproportionate share of the dynamically active structures, scalar-gradient amplification, and heat-release response \citep{HermansonDimotakis1989,DayMansourReynolds2001,PantanoSarkarWilliams2003,KnausPantano2009}. If filtering errors localize in the same subset of the flow, the entropy residual becomes a compact way to quantify where filtering is most damaging. The present analysis addresses that issue using time-resolved mid-plane data from a supersonic reacting shear layer. The main object is the residual
\begin{equation}
\Pi_s^\ast = \overline{\frac{D s^\ast}{D t^\ast}} - \frac{D\overline{s^\ast}}{D t^\ast},
\label{eq:pis}
\end{equation}
which measures the mismatch between the filtered DNS in-plane entropy-transport rate and the same rate reconstructed from filtered fields. The physical interpretation is built through two entropy-production channels: a viscous contribution reconstructed from the full on-plane velocity-gradient tensor and an analogous conductive contribution associated with temperature gradients. These measures do not constitute a complete entropy budget, but they capture the mechanical and thermal pathways most easily accessible from the flow and scalar fields, and which are major contributors to the entropy field.

\section{Configuration and entropy diagnostics}
We analyze DNS data from a three-dimensional supersonic turbulent reacting hydrogen--air temporal mixing layer. The physical configuration follows \citet{OBrien2014}, while the solver-side implementation follows \citet{Oefelein2006-PAS}. The shear layer is sampled on the mid-$z$ plane in time, and all spatio-temporal data are nondimensionalized using the mean convective speed $U_c$ and the initial vorticity thickness $\delta_{\omega,0}$. The entropy-like scalar is
\begin{equation}
s^\ast=\frac{1}{c_{p,\mathrm{ref}}}\left[c_p(Z,T)\ln\!\left(\frac{T}{T_{\mathrm{ref}}}\right)-R_{\mathrm{mix}}(Z,T)\ln\!\left(\frac{\rho}{\rho_{\mathrm{ref}}}\right)\right],
\label{eq:sstar}
\end{equation}
where $c_{p,\mathrm{ref}}$, $T_{\mathrm{ref}}$, and $\rho_{\mathrm{ref}}$ define a fixed reference state, while the local mixture properties vary with the state. This thermodynamic diagnostic retains variable-property temperature and density effects, but not the full species-mixing and chemical-potential contributions. Species-diffusive and chemical kinetic contributions are deliberately not included in this study in order to keep the work concise. For the entropy-like scalar considered here, the material derivative in the mid-plane is given by $\frac{D s^\ast}{D t^\ast}$ with the material derivative operator $\frac{D}{Dt^\ast}$ appropriately nondimensionalized. The specific viscous entropy-production rate and equivalent the conductive contribution are respectively
\begin{equation}
\sigma_\mu^\ast
=
\frac{\delta_{\omega,0}}{U_{c,\mathrm{ref}}c_{p,\mathrm{ref}}}
\frac{1}{\rho}
\frac{\Phi}{T},;\qquad
\sigma_k^\ast
=
\frac{\delta_{\omega,0}}{U_{c,\mathrm{ref}}c_{p,\mathrm{ref}}}
\frac{1}{\rho}
k\frac{|\nabla T|^2}{T^2}.
\label{eq:sigmamu_sigmak}
\end{equation}
with $\Phi=\tau_{ij}\partial_ju_i$. The full on-plane gradient vectors and tensors are reconstructed from stored three-component gradients (computed during the solution procedure), so that the measures are mechanically complete on the sampled plane. 
The same rate scale is used for $\Pi_s^\ast$, so the transport residual and production measures are directly comparable. In \Cref{eq:pis}, the first term is obtained by filtering the DNS in-plane material derivative, while the second is computed from the filtered scalar and filtered in-plane velocity field. Four cases are used: DNS and box-filtered fields $B2$, $B4$, and $B8$, with widths $2\Delta$, $4\Delta$, and $8\Delta$ relative to the grid size $\Delta$.

\section{Results}
\subsection{Instantaneous entropy-transport fields}
\begin{figure}
  \centering
  \includegraphics[width=0.84\textwidth]{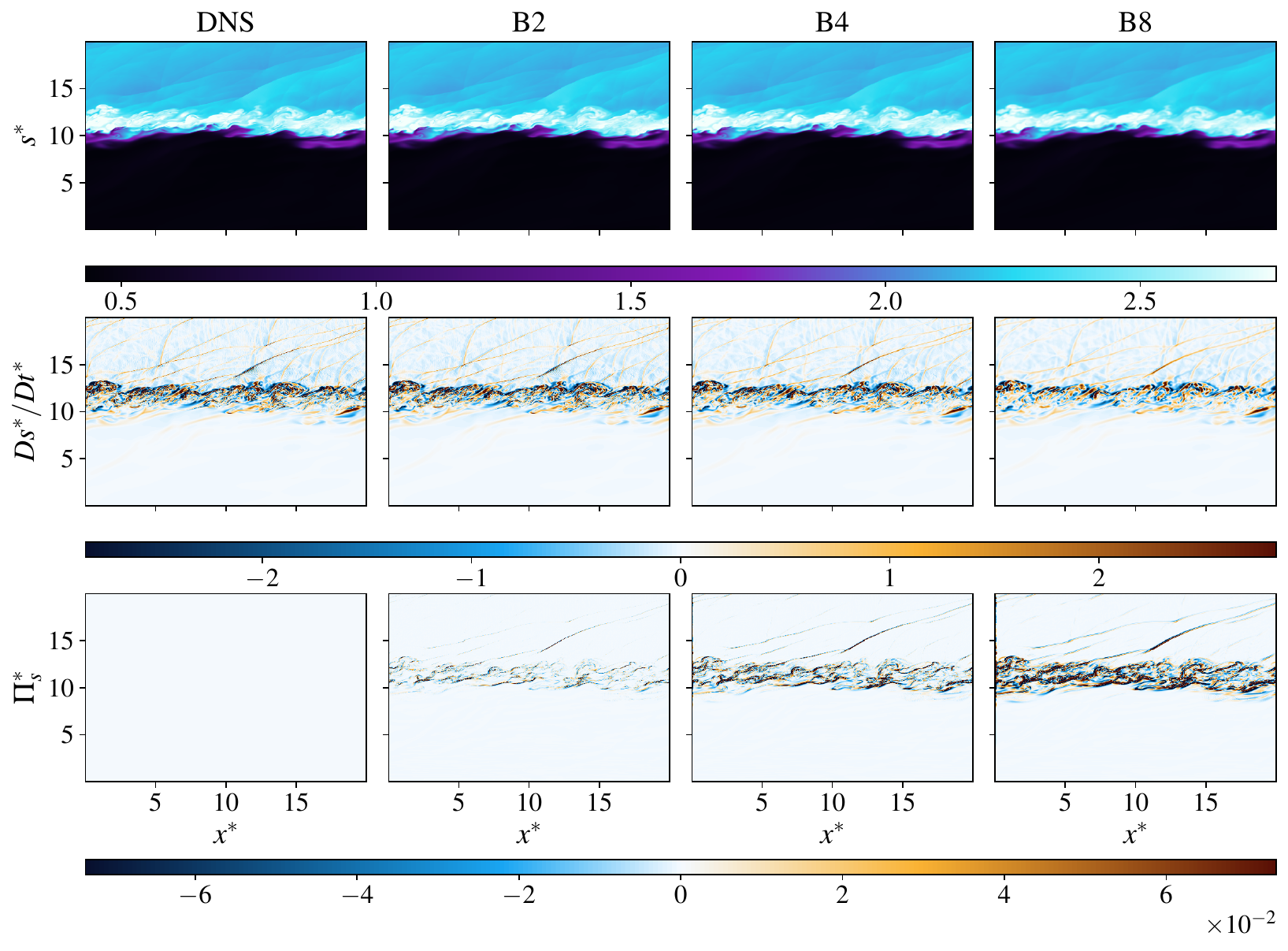}
  \caption{Instantaneous slices of $s^\ast$ (top row), $D s^\ast / D t^\ast$ (middle row), and $\Pi_s^\ast$ (bottom row) for the unfiltered and box-filtered fields.}
  \label{fig:entropy-fields}
\end{figure}
\Cref{fig:entropy-fields} establishes the basic distinction between the entropy field, its transport, and the residual introduced by filtering. The large-scale organization of $s^\ast$ remains almost unchanged from DNS to $B8$, so the primary scalar itself is comparatively robust to the removal of fine scales. By contrast, the most intense structures in $D s^\ast / D t^\ast$ are rapidly attenuated going from $B2$ to $B8$. This difference is the first indication that filtering acts far more strongly on entropy transport than on entropy content. The plane retains the gross cross-stream layering of $s^\ast$, but the derivative field loses the thin structures that carry the largest instantaneous transport activity. In addition, in the supersonic freestream there exist regions that experience sign reversal in $D s^\ast / D t^\ast$ with increasing filter width.

The residual $\Pi_s^\ast$ is more selective still. It does not appear as a weak, domain-wide correction spread across the plane. Instead, it is concentrated in the narrow layer core where the strongest derivative activity was already present in DNS. Its amplitude also grows with filter width. At the representative time shown here, the RMS of $\Pi_s^\ast$ rises from $1.62\times 10^{-2}$ for $B2$ to $4.02\times 10^{-2}$ for $B8$ across the whole plane. Filtering therefore introduces its largest transport discrepancy in the same sharply structured subset of the flow that dominates the unfiltered entropy-rate field, not in the broad outer portions of the layer. In the outer layer, filtering effects are more significant in high-gradient shock regions.

\subsection{Mechanical and thermal entropy-production structure}
\begin{figure}
  \centering
  \includegraphics[width=0.84\textwidth]{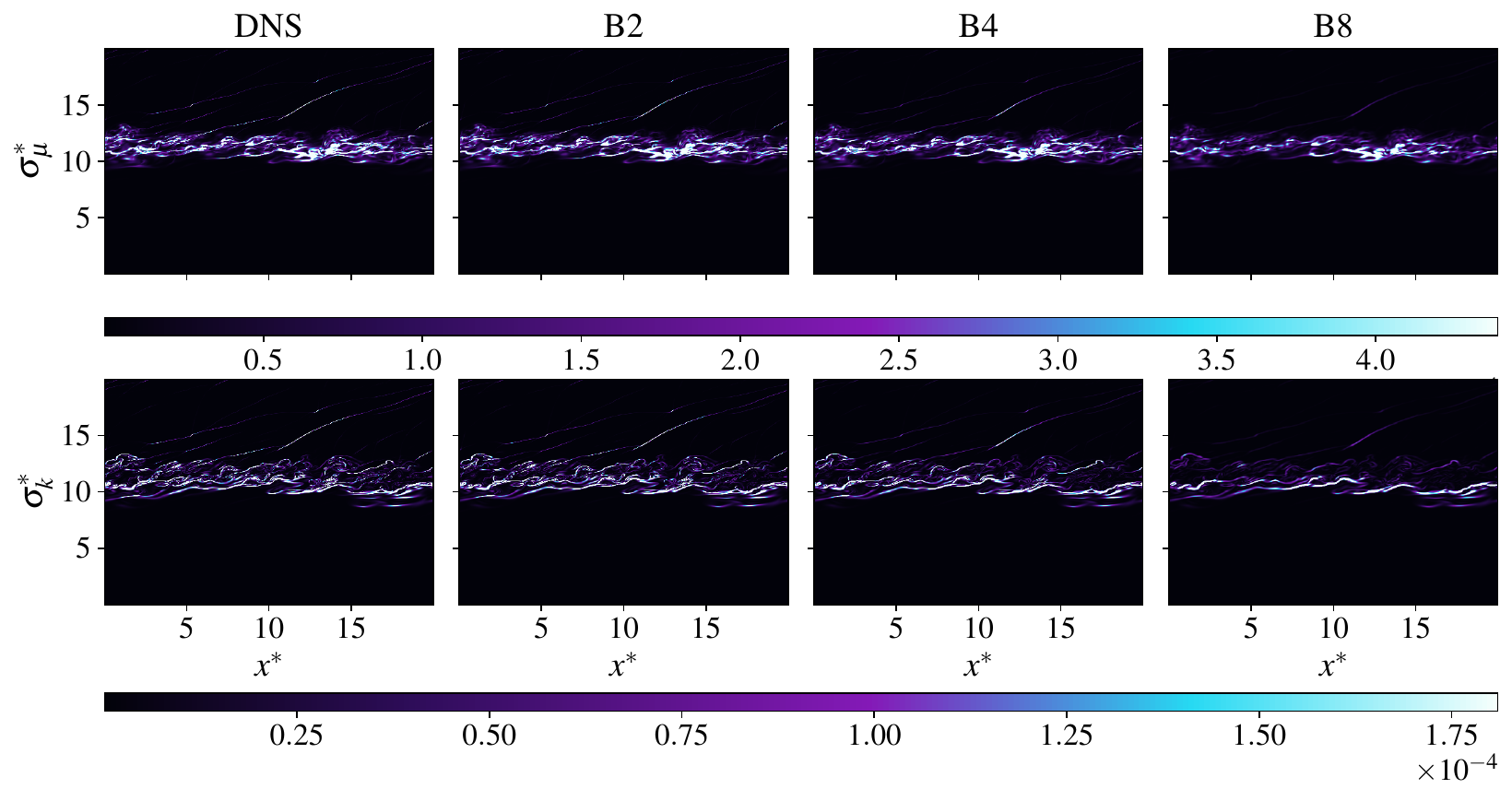}
  \caption{Single-time slices of the viscous and conductive specific entropy-production measures, $\sigma_\mu^\ast$ and $\sigma_k^\ast$.}
  \label{fig:sigma-fields}
\end{figure}

The production fields in \Cref{fig:sigma-fields} show why the residual localizes where it does. The viscous contribution $\sigma_\mu^\ast$ is tightly concentrated in a thin central strip, whereas the conductive contribution $\sigma_k^\ast$ spans a broader thermal layer. That separation is consistent with the established picture of reacting shear layers, in which the most intense strain-driven activity remains concentrated near the layer core while thermal and scalar effects persist across a wider mixed region \citep{HermansonDimotakis1989,PantanoSarkarWilliams2003,MahleEtAl2007}. The thermal/viscous production channels are therefore unique in that they emphasize different but overlapping parts of the same active layer. Beyond the widths of the respective regions being different, the spatial distribution also indicates smaller regions of high magnitude viscous effect in contrast with more distributed but moderate magnitude regions of thermal effect.

Filtering damps both channels, but non-uniformly. At the representative time, the RMS of $\sigma_\mu^\ast$ decreases from $1.29\times 10^{-4}$ for $B2$ to $7.61\times 10^{-5}$ for $B8$, while the RMS of $\sigma_k^\ast$ decreases from $1.00\times 10^{-4}$ to $3.06\times 10^{-5}$ over the same range. The conductive channel is therefore attenuated more aggressively than the viscous channel at that instant, even though both remain centered in the mixed region. The $\Pi_s^\ast$ residual then amplifies as the production channels weaken. In other words, the residual is not merely tracking absolute production amplitude. It is reflecting the increasing mismatch between filtered transport and the unfiltered high-gradient mechanisms that originally sustained that transport.

\subsection{Temporal evolution of the entropy measures}
\begin{figure}
  \centering
  \begin{subfigure}{0.46\textwidth}
    \includegraphics[width=\textwidth]{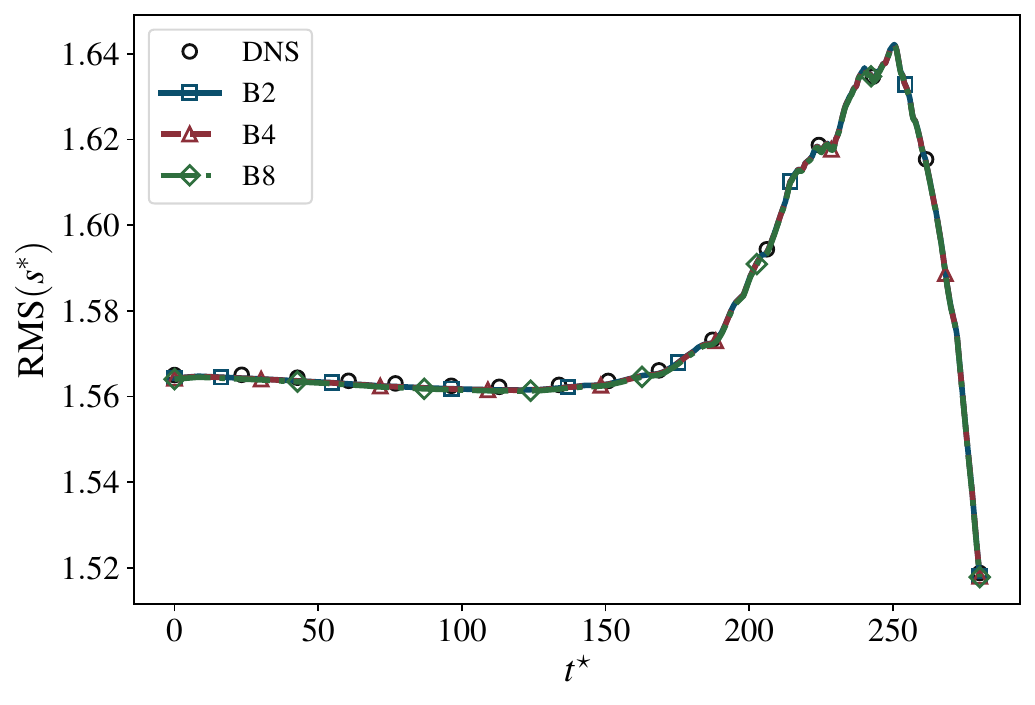}
  \end{subfigure}
  \begin{subfigure}{0.46\textwidth}
    \includegraphics[width=\textwidth]{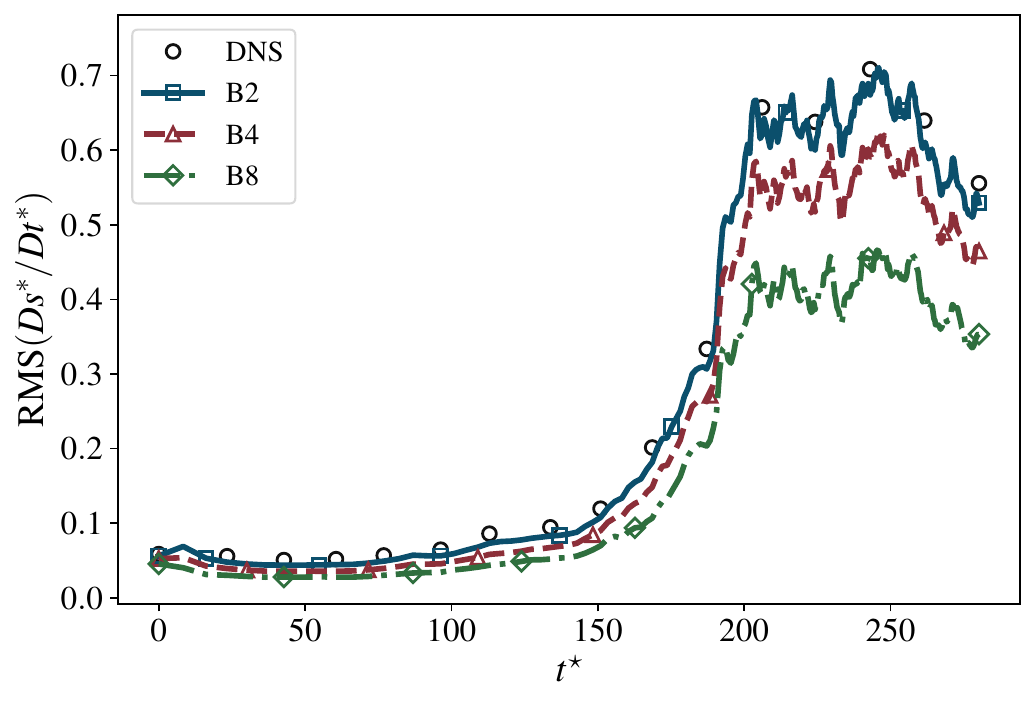}
  \end{subfigure}
  \begin{subfigure}{0.46\textwidth}
    \includegraphics[width=\textwidth]{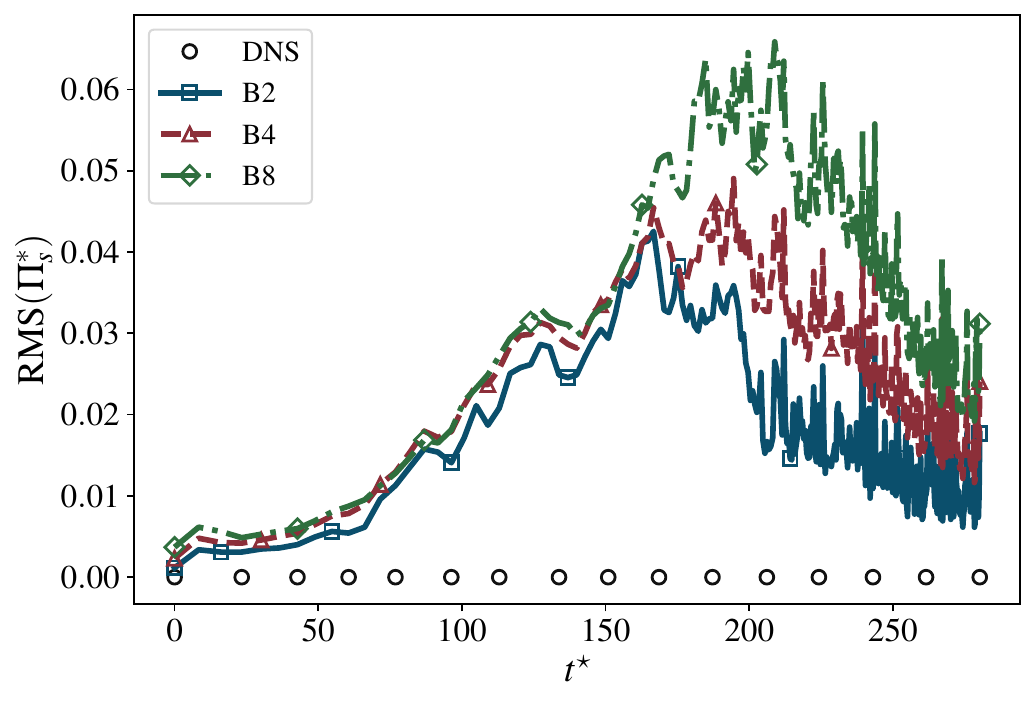}
  \end{subfigure}
  \begin{subfigure}{0.46\textwidth}
    \includegraphics[width=\textwidth]{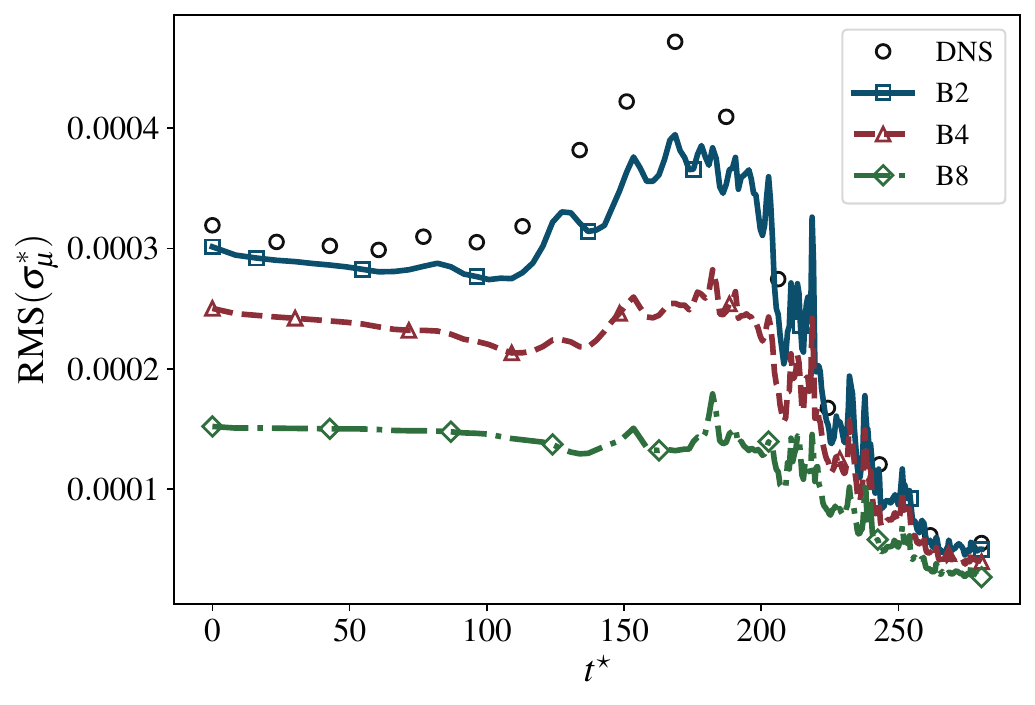}
  \end{subfigure}
  \begin{subfigure}{0.46\textwidth}
    \includegraphics[width=\textwidth]{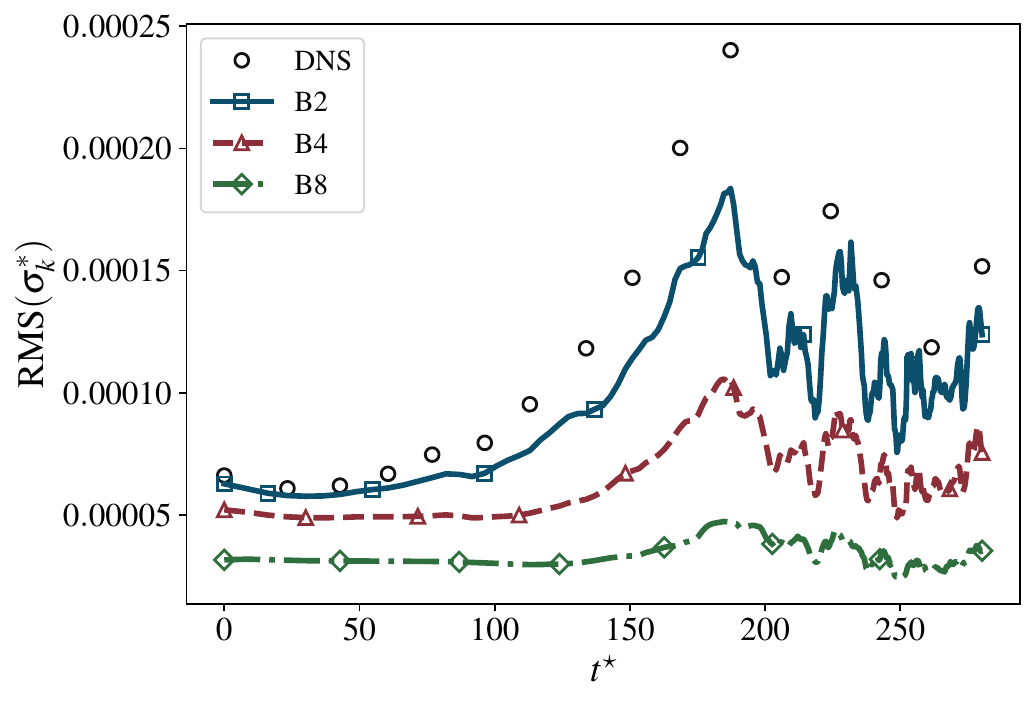}
  \end{subfigure}
  \caption{RMS histories of the main entropy and entropy-production measures.}
  \label{fig:rms}
\end{figure}

The RMS histories in \Cref{fig:rms} preserve the same hierarchy over the full temporal evolution. The entropy field itself is almost insensitive to filtering: the mean RMS of $s^\ast$ changes only from $1.5988$ in DNS to $1.5979$ in $B8$. By contrast, the mean RMS of $D s^\ast / D t^\ast$ decreases from $0.555$ in DNS to $0.344$ in $B8$. The entropy rate is therefore substantially more sensitive to filtering than the entropy field from which it is constructed. This is the temporal analogue of the instantaneous result in \Cref{fig:entropy-fields}: broad field organization survives, while the dynamically sharp transport events do not.

The residual and the two production channels reveal a second layer of organization. The mean RMS of $\Pi_s^\ast$ increases from $1.65\times 10^{-2}$ in $B2$ to $3.67\times 10^{-2}$ in $B8$, whereas the mean RMS of $\sigma_\mu^\ast$ and $\sigma_k^\ast$ decrease by about $56\%$ and $76\%$, respectively, between DNS and $B8$. The peak residual response also shifts toward later stages of the evolution, with the maximum $\Pi_s^\ast$ RMS appearing near $t^\ast\approx 167$ for $B2$, $t^\ast\approx 195$ for $B4$, and $t^\ast\approx 209$ for $B8$. Filtering therefore weakens resolved entropy production while increasing the transport discrepancy that must be carried by the residual. This ordering mirrors the LES expectation that filtering preserves broad field organization more effectively than intense small-scale transport events \citep{Sagaut2006,MeneveauKatz2000,FoysiSarkar2010}, but here the effect is quantified directly in entropy variables. Interestingly, the RMS of the viscous channel peaks and decreases with the temporal evolution, while that of the conductive channel increases and then peaks, indicating that filtering affects these mechanisms on their own different timescales.

\subsection{Distributional response to filtering}
\begin{figure}
  \centering
  \begin{subfigure}{0.46\textwidth}
    \includegraphics[width=\textwidth]{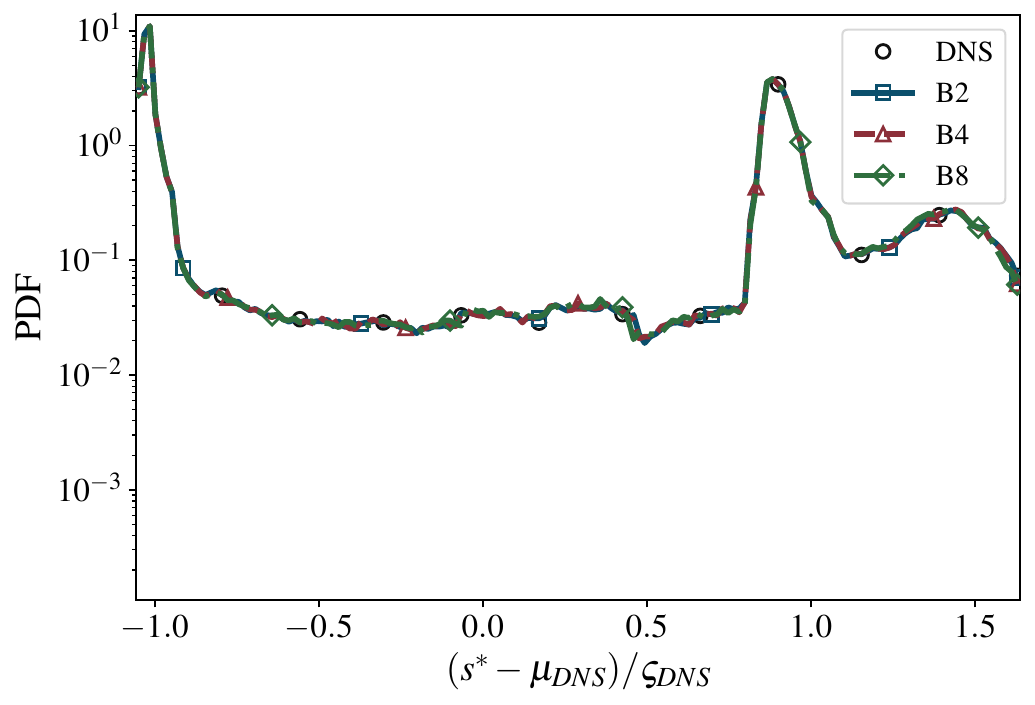}
  \end{subfigure}
  \begin{subfigure}{0.46\textwidth}
    \includegraphics[width=\textwidth]{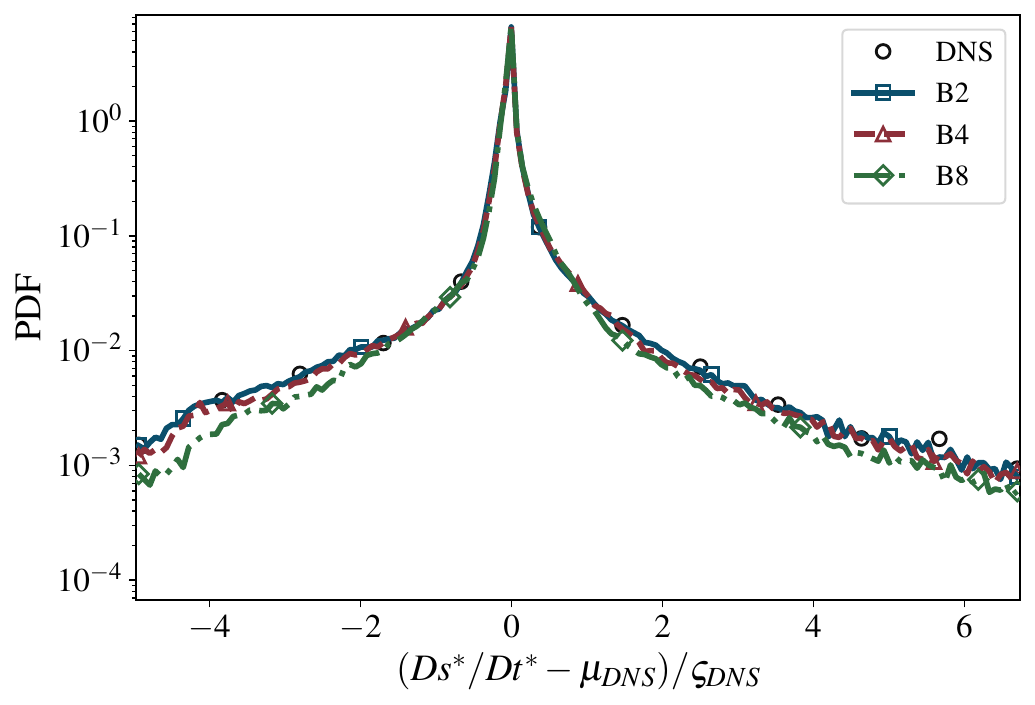}
  \end{subfigure}
  \begin{subfigure}{0.46\textwidth}
    \includegraphics[width=\textwidth]{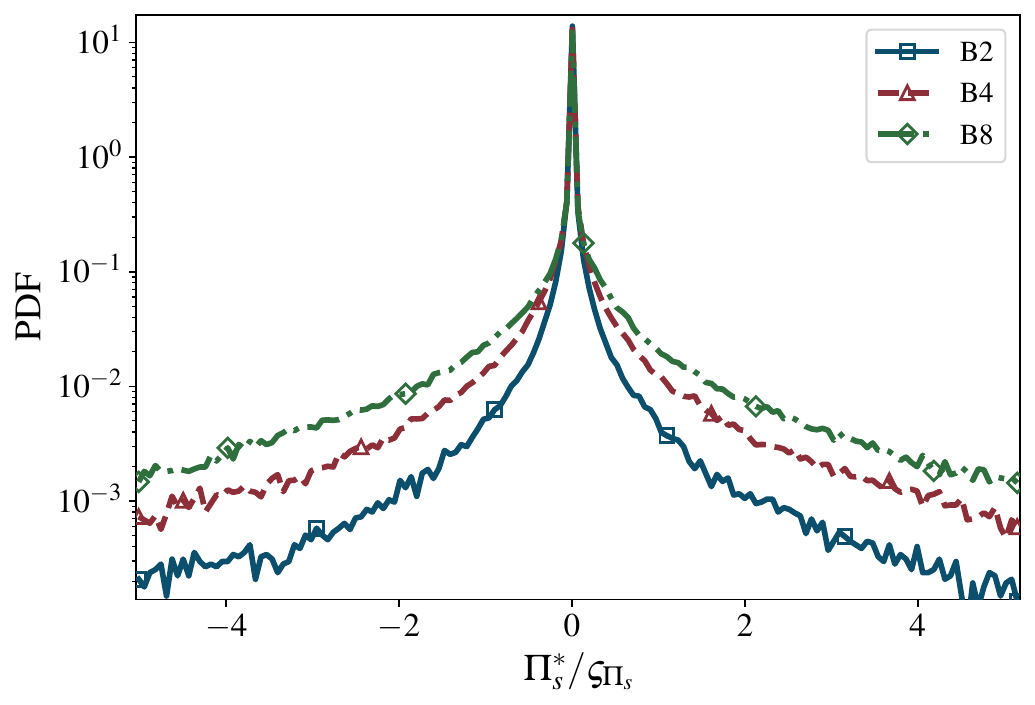}
  \end{subfigure}
  \begin{subfigure}{0.46\textwidth}
    \includegraphics[width=\textwidth]{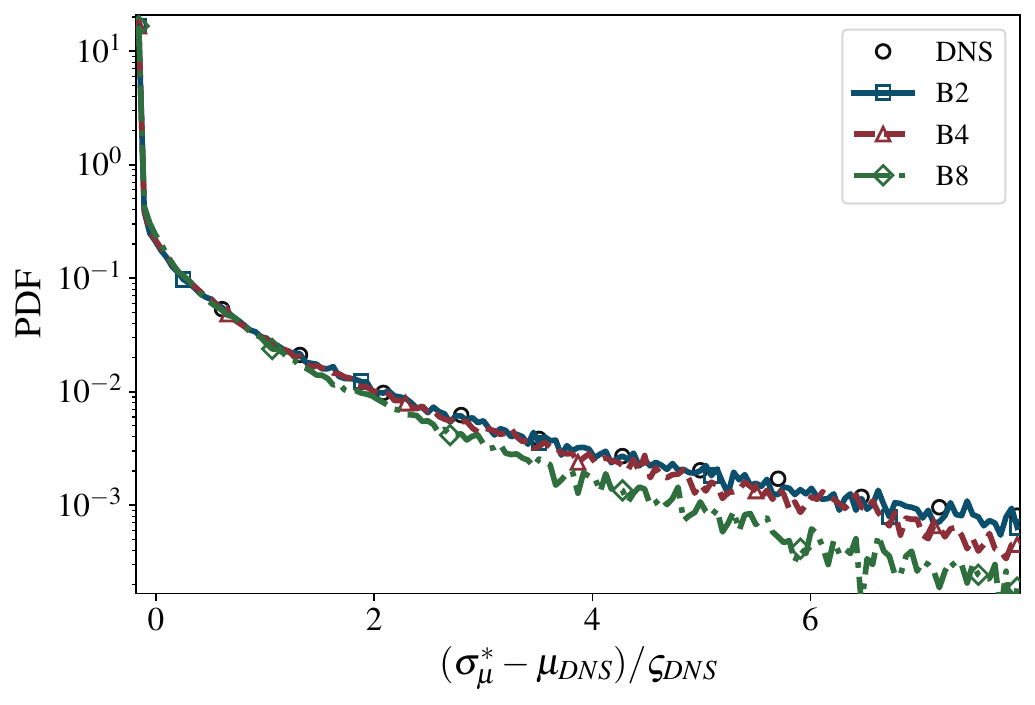}
  \end{subfigure}
  \begin{subfigure}{0.46\textwidth}
    \includegraphics[width=\textwidth]{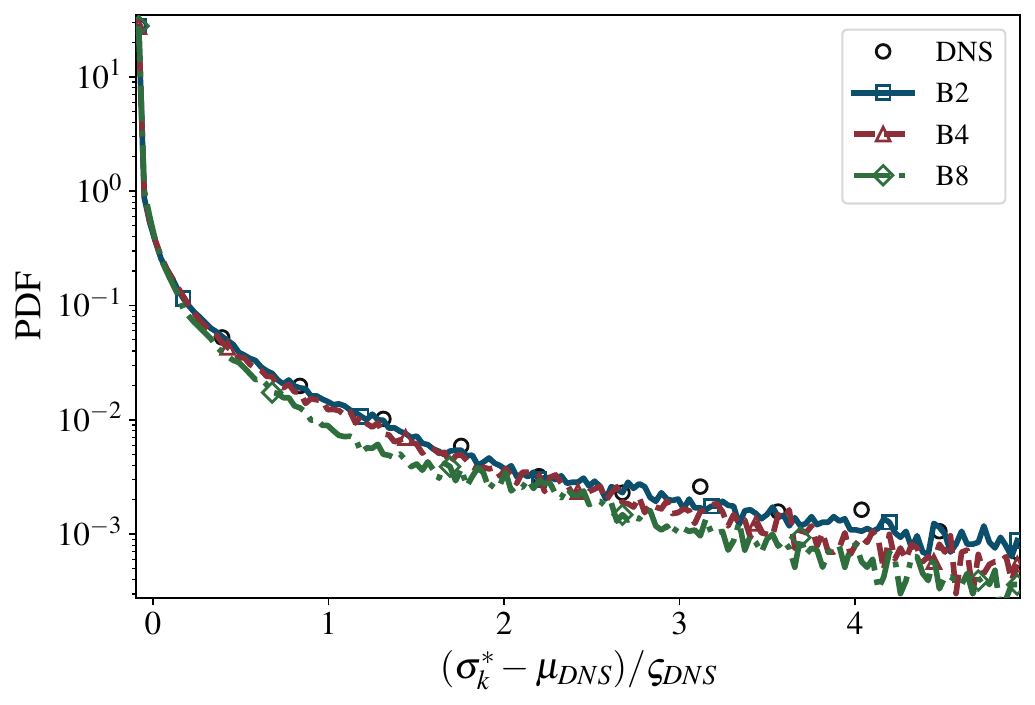}
  \end{subfigure}
  \caption{Probability densities at the same time instant used in \Cref{fig:entropy-fields,fig:sigma-fields}.}
  \label{fig:pdf}
\end{figure}

The PDFs in \Cref{fig:pdf} show that filtering changes the high-amplitude portions of the rate and production fields more strongly than it shifts the central bulk. The distribution of $s^\ast$ changes little. The tails of $D s^\ast / D t^\ast$, $\sigma_\mu^\ast$, and $\sigma_k^\ast$ contract as the filter width increases, indicating attenuation of rare, intense resolved transport and production events. We note that because $\sigma_\mu^\ast$ and $\sigma_k^\ast$ are non-negative, their respective PDFs are one-sided. The residual $\Pi_s^\ast$ behaves differently: its distribution broadens from $B2$ to $B8$ because the filtered-transport mismatch grows with filter width.

This behavior is the statistical counterpart of the field views in \Cref{fig:entropy-fields,fig:sigma-fields}. The strongest resolved transport and production events are the ones least well preserved by coarse filtering, and those events are precisely the ones that populate the far tails of the resolved quantities. The corresponding residual broadening quantifies the transport mismatch that the filtered description must carry. The PDF response therefore confirms that the dominant influence of filtering is focused on the intermittent, sharp, entropy-carrying structures already identified in physical space. This is consistent with prior compressible and reacting shear-layer studies in which the dynamically important gradients and small-scale spectral content occupy only a restricted subset of the mixed region rather than the whole layer \citep{PantanoSarkarWilliams2003,KnausPantano2009,JahanbakhshiMadnia2016}.

\subsection{Residual coupling to production and gradient intensity}
\begin{figure}
  \centering
  \begin{subfigure}{0.47\textwidth}
    \includegraphics[width=\textwidth]{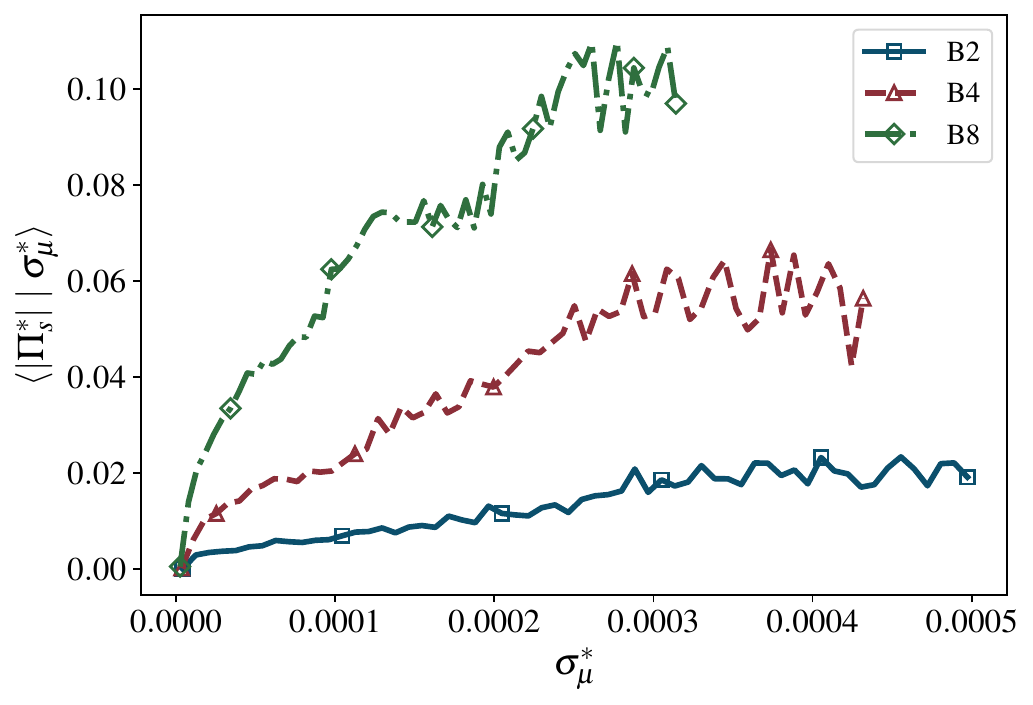}
  \end{subfigure}
  \begin{subfigure}{0.47\textwidth}
    \includegraphics[width=\textwidth]{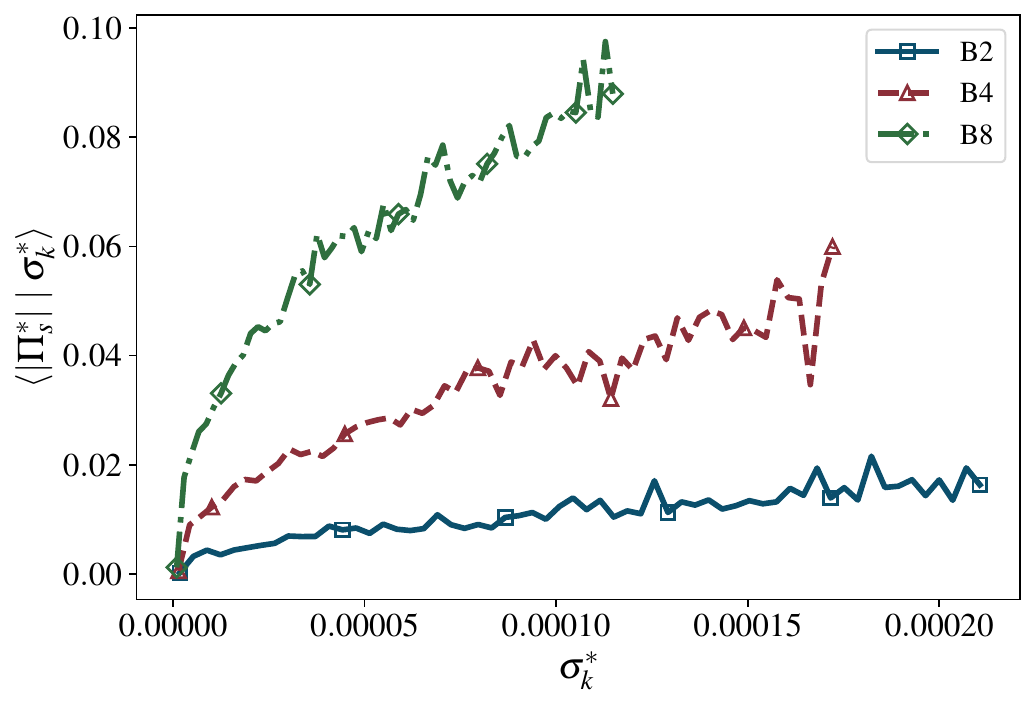}
  \end{subfigure}
  \caption{Conditional mean $|\Pi_s^\ast|$ as a function of the viscous and conductive entropy-production measures.}
  \label{fig:cond-sigma}
\end{figure}

\Cref{fig:cond-sigma} shows that the residual is not merely co-located with the active regions of the layer; it is statistically organized by them. For all three filtered cases, the conditional mean $|\Pi_s^\ast|$ rises monotonically with both $\sigma_\mu^\ast$ and $\sigma_k^\ast$. The increase is also significant. In the representative snapshot, the last conditional bin is larger than the first by factors of roughly $1.9\times 10^2$ for $B2$, $2.3\times 10^2$ for $B4$, and $1.7\times 10^2$ for $B8$ when conditioned on $\sigma_\mu^\ast$. The same trend holds for $\sigma_k^\ast$, although the amplification factors are smaller than for the viscous channel at the same filter width. The residual is therefore most strongly tied to the most intense mechanically active structures, while the conductive channel retains a broader but still systematic influence.

The difference between the viscous and conductive conditionals is physically useful. The viscous channel responds most strongly where the transport field is already thin and sharply organized, so conditioning on $\sigma_\mu^\ast$ produces the steepest rise in $|\Pi_s^\ast|$. The conductive channel still produces a strong monotone response, but the broader thermal footprint softens the growth somewhat. This matches the instantaneous and cross-stream views, and it reinforces the interpretation that the residual is weighted toward the narrowest, most mechanically active subset of the layer rather than being controlled uniformly by all irreversible processes.

\begin{figure}
  \centering
  \begin{subfigure}{0.47\textwidth}
    \includegraphics[width=\textwidth]{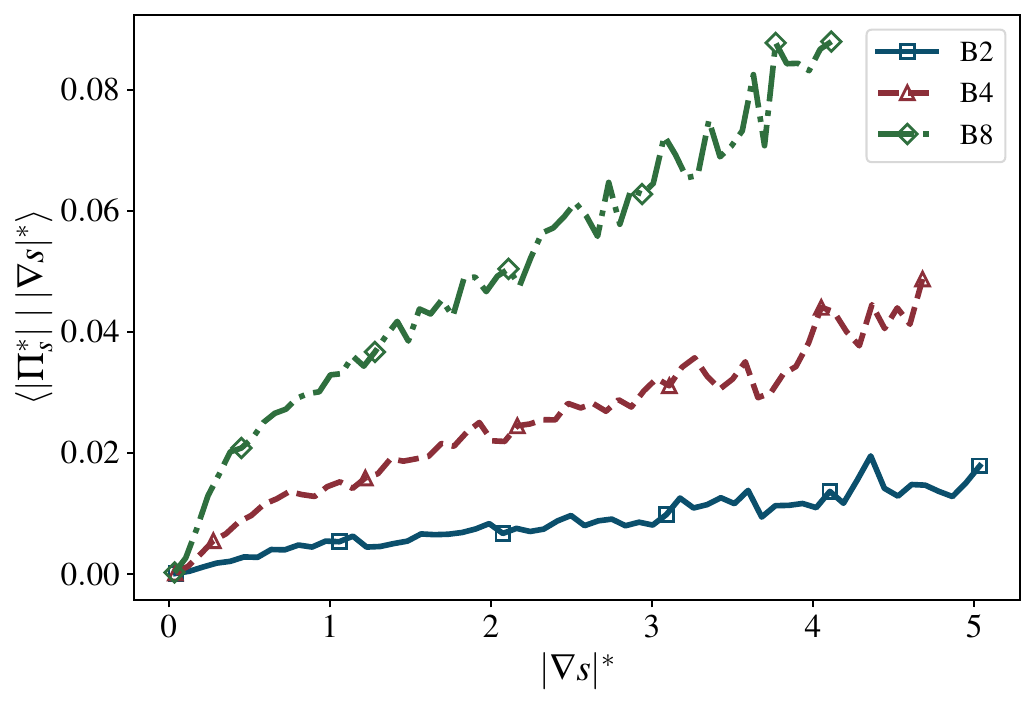}
  \end{subfigure}
  \begin{subfigure}{0.47\textwidth}
    \includegraphics[width=\textwidth]{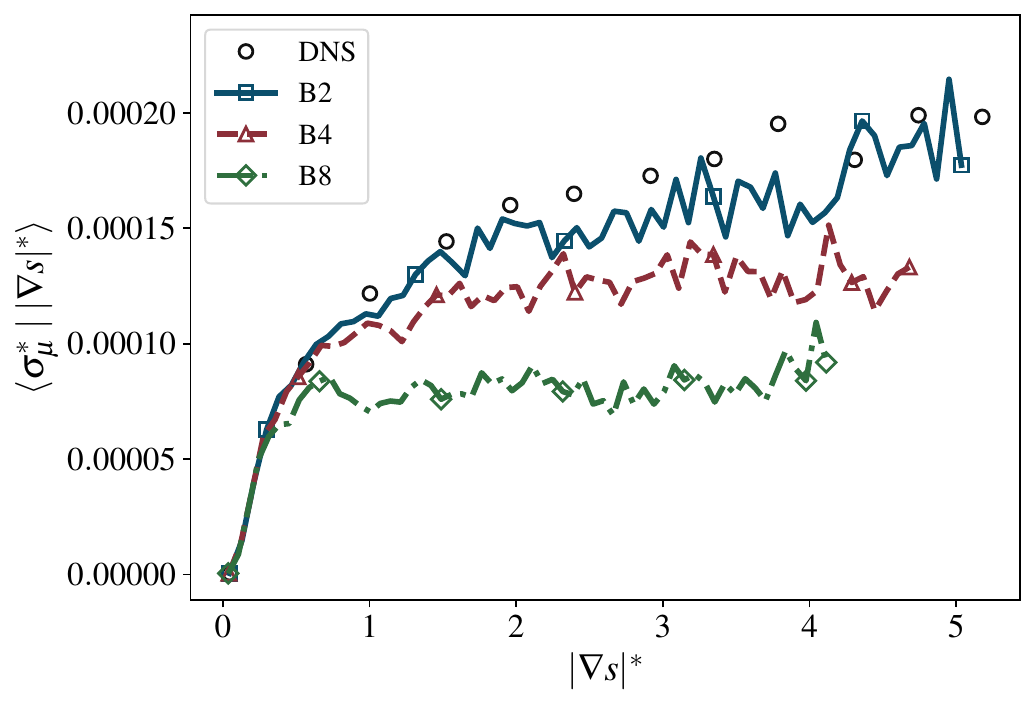}
  \end{subfigure}
  \caption{Conditional relations between entropy-gradient strength and the residual or viscous entropy-production activity.}
  \label{fig:cond-grad}
\end{figure}

The entropy-gradient conditionals in \Cref{fig:cond-grad} also add a further aspect to this dynamic. The residual grows rapidly with $|\nabla s|^\ast$, and the viscous production measure does the same. At the representative time, the conditioned $|\Pi_s^\ast|$ increases by factors of roughly $3.3\times 10^2$, $4.4\times 10^2$, and $4.0\times 10^2$ from the weakest to the strongest gradient bins for $B2$, $B4$, and $B8$, respectively. As expected, the trends are reversed between $|\Pi_s^\ast|$ conditioned on $\sigma_\mu^\ast$ and $\sigma_\mu^\ast$ conditioned on $|\nabla s|^\ast$. For small values of $|\nabla s|^\ast$, there seems to be no significant filter-width effect on the conditional mean for $\sigma_\mu^\ast$.

The residual could have been strongest where filtered transport accumulates weakly over broad areas, but the conditional results show a more localized response. Filtering alters entropy transport most strongly where the layer already contains the steepest entropy gradients and the most intense irreversible activity, which is the same subset of the flow emphasized in prior studies of scalar sharpening and heat-release-modified shear-layer structure \citep{Breidenthal1981,PantanoSarkarWilliams2002,PantanoSarkarWilliams2003}.

\subsection{Spectral and cross-stream organization}
\begin{figure}
  \centering
  \begin{subfigure}{0.47\textwidth}
    \includegraphics[width=\textwidth]{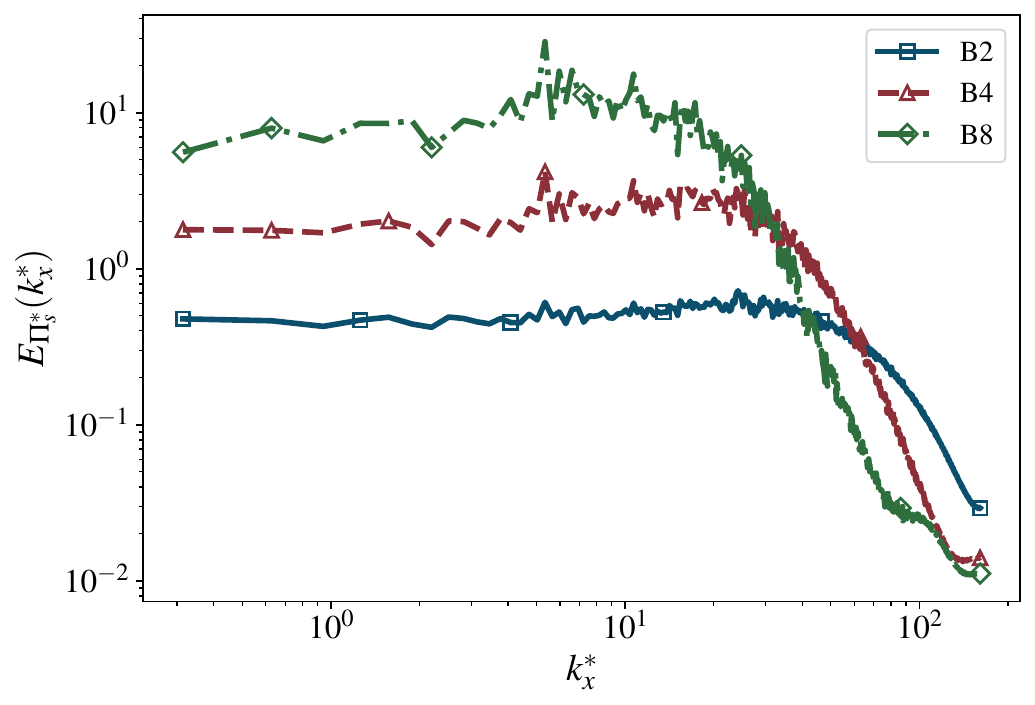}
  \end{subfigure}
  \begin{subfigure}{0.47\textwidth}
    \includegraphics[width=\textwidth]{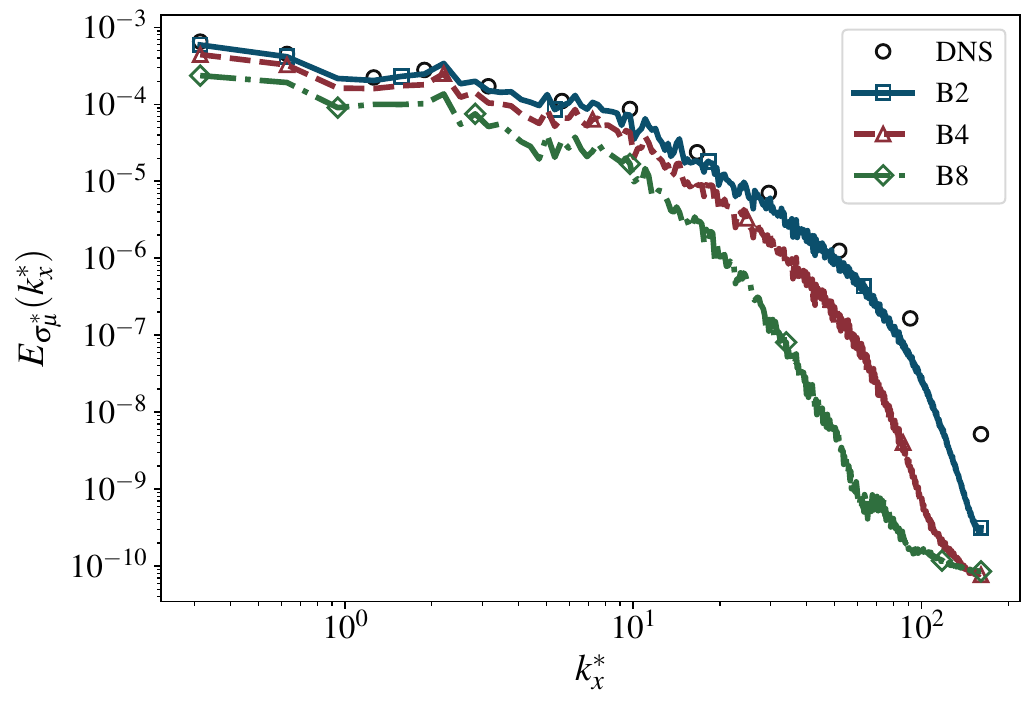}
  \end{subfigure}
  \caption{Streamwise spectra of the residual and the viscous entropy-production measure.}
  \label{fig:spectra}
\end{figure}

\begin{figure}
  \centering
  \begin{subfigure}{0.32\textwidth}
    \includegraphics[width=\textwidth]{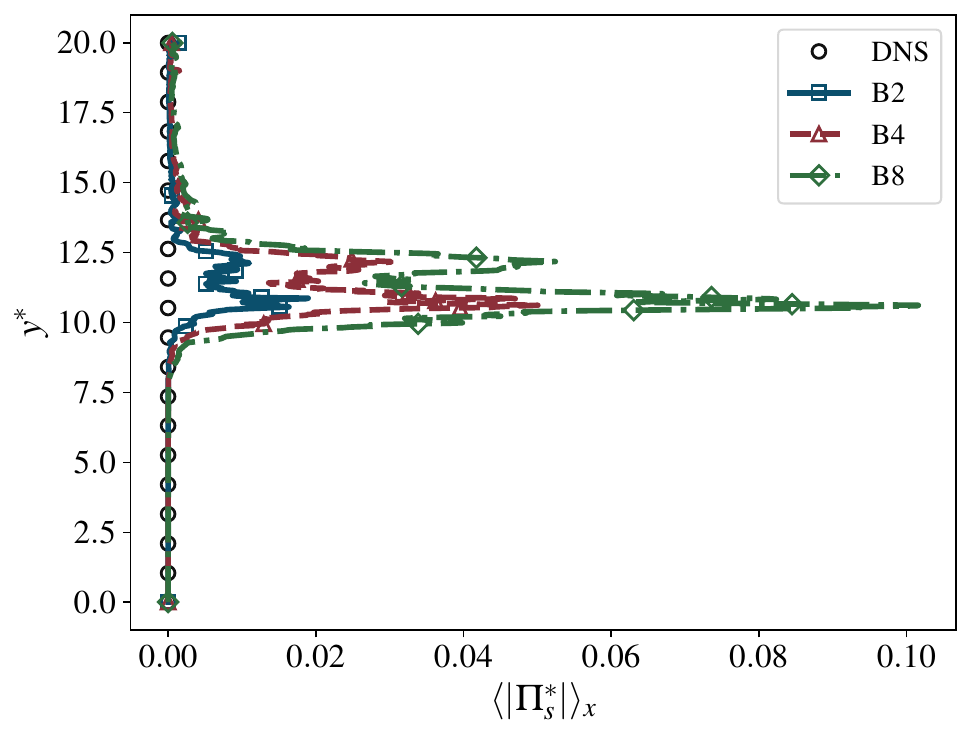}
  \end{subfigure}
  \begin{subfigure}{0.32\textwidth}
    \includegraphics[width=\textwidth]{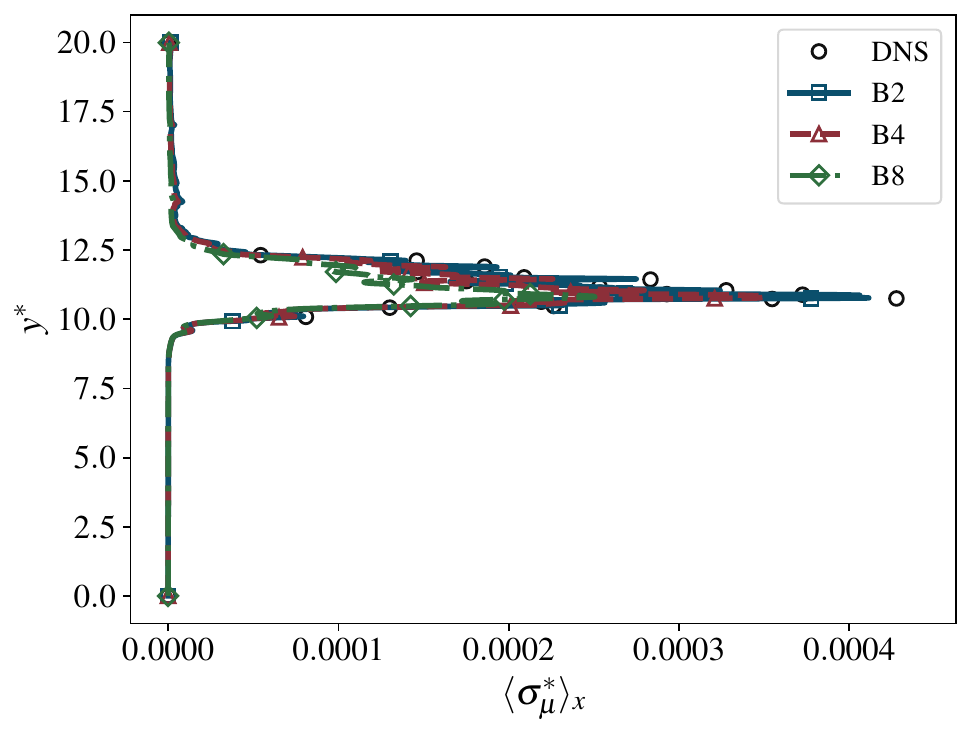}
  \end{subfigure}
  \begin{subfigure}{0.32\textwidth}
    \includegraphics[width=\textwidth]{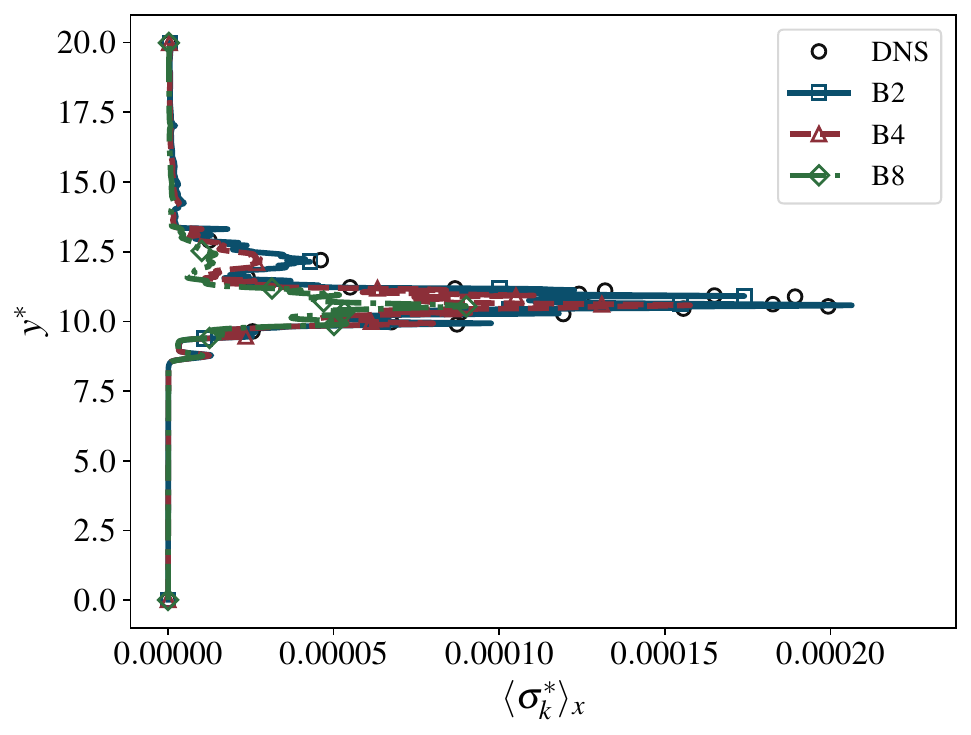}
  \end{subfigure}
  \caption{Cross-stream mean organization of the residual and the two production channels.}
  \label{fig:crossstream}
\end{figure}

The spectral view in \Cref{fig:spectra} shows that filtering acts most strongly on the highest resolved wavenumbers, which is to be expected. In particular, the drop in residual spectral content is sharper for wider filters. For $\Pi_s^\ast$, the high-wavenumber tail fraction in the representative snapshot decreases from $3.9\times 10^{-2}$ for $B2$ to $1.9\times 10^{-3}$ for $B8$. The corresponding fraction for $\sigma_\mu^\ast$ is already much smaller in DNS and decreases further under filtering. In conjunction with the PDF trends, this shows that $\Pi_s^\ast$ broadens in amplitude space while losing high-$k_x^\ast$ content in spectral space. The result also aligns with previous observations that reacting and compressible shear layers carry a substantial part of their dynamically relevant structure in narrow high-gradient bands and small-scale spectral content \citep{PantanoSarkar2002,KnausPantano2009,MahleEtAl2007}.

The cross-stream profiles (averaged along the homogeneous directions) in \Cref{fig:crossstream} show where that attenuation remains anchored. The largest values of $\langle |\Pi_s^\ast| \rangle_x$, $\langle \sigma_\mu^\ast \rangle_x$, and $\langle \sigma_k^\ast \rangle_x$ all remain centered in the mixed region rather than migrating into the outer freestreams. The conductive channel spans a slightly broader cross-stream extent than the viscous channel, but both remain organized by the same core subset in which the residual is strongest. Spectral attenuation and cross-stream localization therefore tell a consistent story, that filtering removes the finest content first, but the surviving discrepancies remain concentrated in the same layer core rather than diffusing outward into dynamically weak regions, which is consistent with the mean-value properties of the box filter.

\subsection{Structural simplification of high-residual regions}
\begin{figure}
  \centering
  \begin{subfigure}{0.46\textwidth}
    \includegraphics[width=\textwidth]{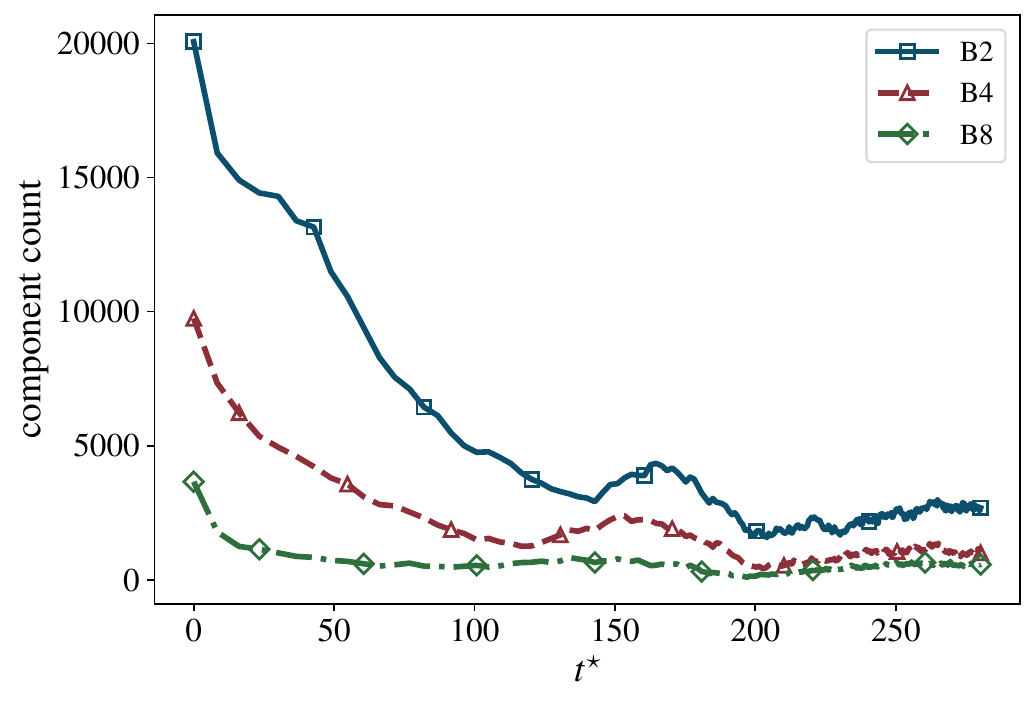}
  \end{subfigure}
  \begin{subfigure}{0.46\textwidth}
    \includegraphics[width=\textwidth]{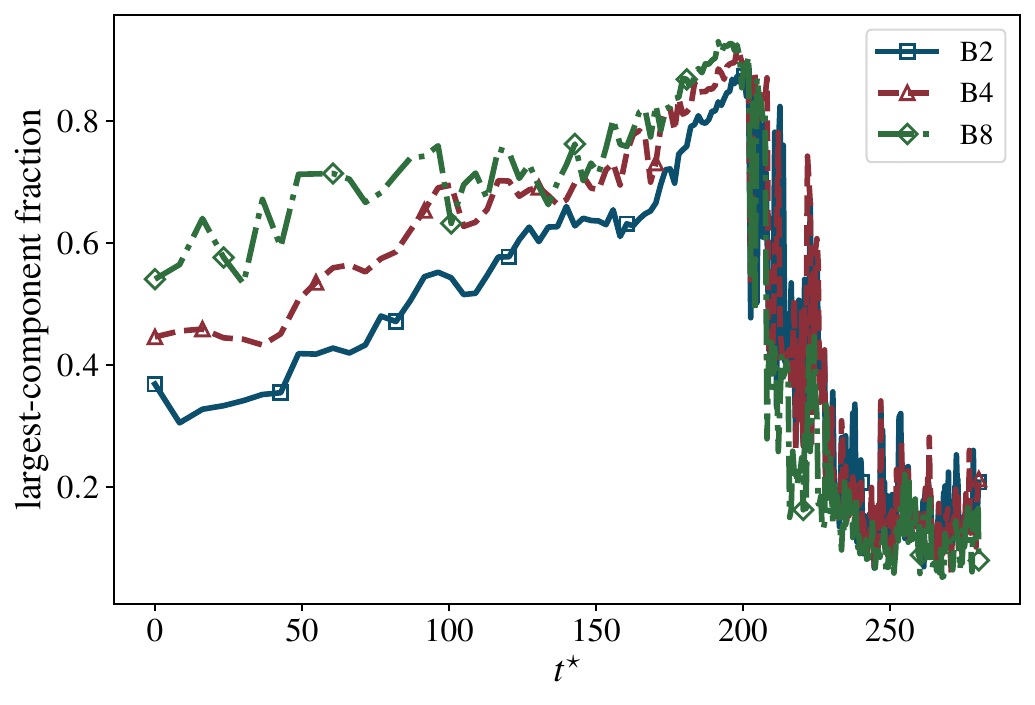}
  \end{subfigure}
  \caption{Connected-component statistics for the high-$|\Pi_s^\ast|$ sets defined by the $95$th percentile threshold.}
  \label{fig:structure}
\end{figure}

Filtering changes not only the magnitude of the residual but also the geometry of the regions that carry the largest residual values. In \Cref{fig:structure}, the number of connected components in the high-$|\Pi_s^\ast|$ sets decreases strongly with filter width. Over the full record, the mean component count decreases from about $3.0\times 10^3$ for $B2$ to about $5.2\times 10^2$ for $B8$. The reduction is even more dramatic at the earliest times, where the count drops from $2.01\times 10^4$ to $3.66\times 10^3$. The strongest residual regions therefore become progressively less fragmented as the filter width increases.

The largest-component fraction remains strongly time dependent, which indicates that the residual geometry is still modulated by the evolving layer dynamics and not determined by filter width alone. Even so, the broad trend is consistent with geometric simplification as the finest corrugations are removed. This is the structural counterpart of the spectral attenuation in \Cref{fig:spectra}, that once the highest-wavenumber corrugations are lost, the residual is carried by a smaller number of broader coherent sets. Filtering therefore reorganizes the entropy-transport error in both amplitude and geometry.

\section{Conclusions}
Filtering alters entropy transport in the reacting shear layer much more strongly than it alters the entropy field itself. The entropy-like scalar $s^\ast$ is comparatively robust, whereas the transport rate $D s^\ast / D t^\ast$ (in particular, the gradient $|\nabla s|^\ast$) and the specific entropy-production measures $\sigma_\mu^\ast$ and $\sigma_k^\ast$ are all strongly attenuated by increasing filter width. At the same time, the residual $\Pi_s^\ast$ grows and remains concentrated in the same narrow layer core that carries the strongest mechanical and thermal entropy-production activity.

Three points emerge consistently across the field, statistical, spectral, and structural views. First, filtering distorts entropy transport preferentially in the most active part of the layer rather than in the outer weak-gradient regions. In the outer supersonic freestream, filtering affects the shocks more preferentially. Second, the residual is organized by both entropy-production intensity and entropy-gradient strength, which ties the filtering error directly to the sharp structures that dominate the core. Third, increasing filter width removes the highest-wavenumber content and simplifies the geometry of the high-$|\Pi_s^\ast|$ sets. The main consequence of filtering is therefore a non-uniform smoothing of the entropy transport field, along with a targeted reorganization of entropy transport and its viscous/conductive mechanisms through the mechanically and thermally active structures that dominate the reacting shear layer.

\begin{acknowledgments}
\noindent\textbf{Acknowledgments.}
The computing power for this study was provided by the Phoenix Computing Cluster as a part of Georgia Tech's Partnership for Advanced Computing Environment, and is gratefully acknowledged.
\end{acknowledgments}

\noindent\textbf{Funding.}
This research received no specific grant from any funding agency, commercial or not-for-profit sectors.

\noindent\textbf{Author contributions.}
Sriram P. Kalathoor: Conceptualization, Methodology, Software, Formal analysis, Investigation, Visualization, Writing -- original draft. Joseph C. Oefelein: Conceptualization, Resources, Supervision, Writing -- review and editing.

\noindent\textbf{Declaration of interests.}
The authors report no conflict of interest.

\noindent\textbf{Data availability statement.}
The data that support the findings of this study are available from the corresponding author upon reasonable request.

\bibliographystyle{jfm}
\bibliography{references}

@article{Oefelein2006-PAS,
  author  = {Oefelein, Joseph C.},
  title   = {Large eddy simulation of turbulent combustion processes in propulsion and power systems},
  journal = {Progress in Aerospace Sciences},
  volume  = {42},
  number  = {1},
  pages   = {2--37},
  year    = {2006},
  issn    = {0376-0421},
  doi     = {10.1016/j.paerosci.2006.02.001}
}

@article{OBrien2014,
  author  = {O'Brien, J. and Urzay, J. and Ihme, M. and Moin, P. and Saghafian, A.},
  title   = {Subgrid-scale backscatter in reacting and inert supersonic hydrogen-air turbulent mixing layers},
  journal = {Journal of Fluid Mechanics},
  volume  = {743},
  pages   = {554--584},
  year    = {2014},
  doi     = {10.1017/jfm.2014.62}
}

@book{DeGrootMazur1962,
  author = {{de Groot}, S. R. and Mazur, P.},
  title = {Non-Equilibrium Thermodynamics},
  publisher = {North-Holland},
  address = {Amsterdam},
  year = {1962}
}

@article{BrownRoshko1974,
  author = {Brown, G. L. and Roshko, A.},
  title = {On density effects and large structure in turbulent mixing layers},
  journal = {Journal of Fluid Mechanics},
  volume = {64},
  number = {4},
  pages = {775--816},
  year = {1974},
  doi = {10.1017/S002211207400190X}
}

@article{Breidenthal1981,
  author = {Breidenthal, Robert E.},
  title = {Structure in turbulent mixing layers and wakes using a chemical reaction},
  journal = {Journal of Fluid Mechanics},
  volume = {109},
  pages = {1--24},
  year = {1981},
  doi = {10.1017/S0022112081000906}
}

@article{BroadwellBreidenthal1982,
  author = {Broadwell, James E. and Breidenthal, Robert E.},
  title = {A simple model of mixing and chemical reaction in a turbulent shear layer},
  journal = {Journal of Fluid Mechanics},
  volume = {125},
  pages = {397--410},
  year = {1982},
  doi = {10.1017/S0022112082003401}
}

@article{PapamoschouRoshko1988,
  author = {Papamoschou, D. and Roshko, A.},
  title = {The compressible turbulent shear layer: an experimental study},
  journal = {Journal of Fluid Mechanics},
  volume = {197},
  pages = {453--477},
  year = {1988},
  doi = {10.1017/S0022112088003325}
}

@article{SandhamReynolds1991,
  author = {Sandham, N. D. and Reynolds, W. C.},
  title = {Three-dimensional simulations of large eddies in the compressible mixing layer},
  journal = {Journal of Fluid Mechanics},
  volume = {224},
  pages = {133--158},
  year = {1991},
  doi = {10.1017/S0022112091001684}
}

@article{HermansonDimotakis1989,
  author = {Hermanson, J. C. and Dimotakis, P. E.},
  title = {Effects of heat release in a turbulent, reacting shear layer},
  journal = {Journal of Fluid Mechanics},
  volume = {199},
  pages = {333--375},
  year = {1989},
  doi = {10.1017/S0022112089000406}
}

@article{DayMansourReynolds2001,
  author = {Day, M. J. and Mansour, N. N. and Reynolds, W. C.},
  title = {Nonlinear stability and structure of compressible reacting mixing layers},
  journal = {Journal of Fluid Mechanics},
  volume = {446},
  pages = {375--408},
  year = {2001},
  doi = {10.1017/S002211200100595X}
}

@article{PantanoSarkar2002,
  author = {Pantano, Carlos and Sarkar, Sutanu},
  title = {A study of compressibility effects in the high-speed turbulent shear layer using direct simulation},
  journal = {Journal of Fluid Mechanics},
  volume = {451},
  pages = {329--371},
  year = {2002},
  doi = {10.1017/S0022112001006978}
}

@article{PantanoSarkarWilliams2003,
  author = {Pantano, Carlos and Sarkar, Sutanu and Williams, Forman A.},
  title = {Mixing of a conserved scalar in a turbulent reacting shear layer},
  journal = {Journal of Fluid Mechanics},
  volume = {481},
  pages = {291--328},
  year = {2003},
  doi = {10.1017/S0022112003003872}
}

@article{LivescuJaberiMadnia2002,
  author = {Livescu, Daniel and Jaberi, Fereshteh A. and Madnia, Cyrus K.},
  title = {The effects of heat release on the energy exchange in turbulent reacting shear flow},
  journal = {Journal of Fluid Mechanics},
  volume = {450},
  pages = {35--66},
  year = {2002},
  doi = {10.1017/S0022112001006164}
}

@incollection{PantanoSarkarWilliams2002,
  author = {Pantano, Carlos and Sarkar, Sutanu and Williams, Forman A.},
  title = {The interaction of scalar mixing and heat release in a reacting shear layer},
  booktitle = {IUTAM Symposium on Turbulent Mixing and Combustion},
  editor = {Pollard, A. and Candel, S.},
  series = {Fluid Mechanics and Its Applications},
  volume = {70},
  publisher = {Springer},
  address = {Dordrecht},
  year = {2002},
  pages = {137--148},
  doi = {10.1007/978-94-017-1998-8_11}
}

@article{MahleEtAl2007,
  author = {Mahle, Inga and Foysi, Holger and Sarkar, Sutanu and Friedrich, Rainer},
  title = {On the turbulence structure in inert and reacting compressible mixing layers},
  journal = {Journal of Fluid Mechanics},
  volume = {593},
  pages = {171--180},
  year = {2007},
  doi = {10.1017/S0022112007008919}
}

@article{KnausPantano2009,
  author = {Knaus, R. C. and Pantano, C.},
  title = {On the effect of heat release in turbulence spectra of non-premixed reacting shear layers},
  journal = {Journal of Fluid Mechanics},
  volume = {626},
  pages = {67--109},
  year = {2009},
  doi = {10.1017/S0022112008005636}
}

@article{FoysiSarkar2010,
  author = {Foysi, H. and Sarkar, S.},
  title = {The compressible mixing layer: an {LES} study},
  journal = {Theoretical and Computational Fluid Dynamics},
  volume = {24},
  number = {6},
  pages = {565--588},
  year = {2010},
  doi = {10.1007/s00162-009-0176-8}
}

@article{JahanbakhshiMadnia2016,
  author = {Jahanbakhshi, R. and Madnia, C. K.},
  title = {Entrainment in a compressible turbulent shear layer},
  journal = {Journal of Fluid Mechanics},
  volume = {797},
  pages = {564--603},
  year = {2016},
  doi = {10.1017/jfm.2016.296}
}

@article{JahanbakhshiMadnia2018,
  author = {Jahanbakhshi, R. and Madnia, C. K.},
  title = {The effect of heat release on the entrainment in a turbulent mixing layer},
  journal = {Journal of Fluid Mechanics},
  volume = {844},
  pages = {92--126},
  year = {2018},
  doi = {10.1017/jfm.2018.122}
}

@book{Pope2000,
  author = {Pope, Stephen B.},
  title = {Turbulent Flows},
  publisher = {Cambridge University Press},
  address = {Cambridge},
  year = {2000}
}

@book{Sagaut2006,
  author = {Sagaut, Pierre},
  title = {Large Eddy Simulation for Incompressible Flows: An Introduction},
  edition = {3rd},
  publisher = {Springer},
  address = {Berlin},
  year = {2006}
}

@article{MeneveauKatz2000,
  author = {Meneveau, Charles and Katz, Joseph},
  title = {Scale-invariance and turbulence models for large-eddy simulation},
  journal = {Annual Review of Fluid Mechanics},
  volume = {32},
  pages = {1--32},
  year = {2000},
  doi = {10.1146/annurev.fluid.32.1.1}
}

@article{Lele1994,
  author = {Lele, Sanjiva K.},
  title = {Compressibility effects on turbulence},
  journal = {Annual Review of Fluid Mechanics},
  volume = {26},
  pages = {211--254},
  year = {1994},
  doi = {10.1146/annurev.fl.26.010194.001235}
}

\end{document}